\documentstyle[psfig,conf_iap]{article}
\begin{document}
\newcommand{\hmpc}{h^{-1}\;{\rm Mpc}}
\newcommand{\lya}{Ly$\alpha$\ }
\newcommand{\overden}{{\rho/{\overline\rho}}}
\newcommand{\K}{{\rm K}}
\newcommand{\kms}{{\rm km}\;{\rm s}^{-1}}
\newcommand{\Nh}{N_{\rm HI}}
\newcommand{\nh}{n_{\rm HI}}
\newcommand{\cdunits}{{\rm cm}^{-2}}
\newbox\grsign \setbox\grsign=\hbox{$>$} \newdimen\grdimen \grdimen=\ht\grsign
\newbox\simlessbox \newbox\simgreatbox
\setbox\simgreatbox=\hbox{\raise.5ex\hbox{$>$}\llap
     {\lower.5ex\hbox{$\sim$}}}\ht1=\grdimen\dp1=0pt
\setbox\simlessbox=\hbox{\raise.5ex\hbox{$<$}\llap
     {\lower.5ex\hbox{$\sim$}}}\ht2=\grdimen\dp2=0pt
\newcommand{\simgt}{\mathrel{\copy\simgreatbox}}
\newcommand{\simlt}{\mathrel{\copy\simlessbox}}
\newcommand{\ga}{\mathrel{\copy\simgreatbox}}
\newcommand{\la}{\mathrel{\copy\simlessbox}}

\heading{%
%
Hubble Flow Broadening of the Lyman-alpha Forest and its 
Implications\footnote{To appear in Proceedings of the 13th IAP Colloquium,
Structure and Evolution of the Intergalactic Medium from QSO
Absorption Line Systems, eds. P. Petitjean \& S. Charlot,
(Nouvelles Fronti\`eres: Paris)}
%
} 
\par\medskip\noindent
\author{%
David H. Weinberg$^1$, Lars Hernquist$^2$, Neal Katz$^3$, 
Rupert Croft$^1$, Jordi Miralda-Escud\'e$^4$
}
\address{
Ohio State University, Dept.\ of Astronomy, Columbus, OH 43210
}
\address{
University of California, Lick Observatory, Santa Cruz, CA 95064
}
\address{
University of Massachusetts, Dept.\ of Physics and Astronomy, Amherst, MA, 01003
}
\address{
University of Pennsylvania,
Dept.\ of Physics and Astronomy, Philadelphia, PA 19104
}

\begin{abstract}
\lya forest lines in QSO spectra have typical widths of $20-50\;\kms$.
Low column density absorbers in cosmological simulations are large,
diffuse structures, and the Hubble flow across the spatially extended
absorber is usually the dominant contribution to the width of its associated 
absorption line.  Thermal broadening is unimportant over most of the spectrum, 
and peculiar velocities tend to make absorption features narrower rather
than broader.  As a consequence of Hubble flow broadening, there is
a close relation between local \lya optical depth and local neutral
hydrogen density, which is well approximated by the Gunn-Peterson formula.
The physics that governs the unshocked intergalactic medium leads to 
a tight correlation between the neutral hydrogen density and the underlying
gas and dark matter overdensity.  For many purposes, it is simpler to
regard a \lya forest spectrum as a continuous, non-linear map of the 
density field rather than a collection of discrete lines.  
This continuous field view of the \lya forest can be applied to measurement 
of the baryon density parameter, testing of cosmological models, and robust 
determination of the shape and amplitude of the primordial mass power spectrum.
\end{abstract}

\medskip

In the last few years, the study of the \lya forest has undergone
several observational revolutions: the extension to low redshift
via HST, the probe of internal structure from spectra along
neighboring lines of sight, the extraordinary detail 
provided by Keck HIRES data, and the clear detections
of metals associated with low column density HI absorbers.
The field has also undergone a theoretical revolution, driven by
hydrodynamic cosmological simulations.
These allow one to predict properties of the \lya forest from 
{\it a priori} theoretical models motivated by independent considerations
of large scale structure, the cosmic microwave background, and
galaxy formation.  
Since the pioneering numerical study of Cen et al.\ \cite{cen94},
there have been more than 30 papers using
cosmological simulations to investigate QSO absorption phenomena, 
by several independent groups.
The resulting picture of the \lya forest
has features in common with some earlier models, especially the
fluctuating intergalactic medium (IGM) scenario of Bi \cite{bi93}.

One distinctive feature of this cosmological picture of the \lya forest
is the low density of the absorbing structures.
Typical marginally saturated lines ($\Nh \sim 10^{14}\;\cdunits$)
arise in gas whose density is a few times the cosmic mean or less.
Weak lines ($\Nh \simlt 10^{13}\;\cdunits$) often occur at local
maxima that lie below the global mean density.
This low density has a number of important consequences.
Most absorption arises in structures that are still expanding 
with residual Hubble flow.
These absorbing systems are usually far from dynamical, hydrostatic,
or thermal equilibrium.
The low density implies a low recombination rate and thus a low
neutral fraction (typically $\sim 10^{-6} - 10^{-4}$),
so the neutral hydrogen revealed by the observed \lya opacity
is only the tip of a much larger iceberg.

The low density of the absorbing gas also means that absorbers must
be physically large in order to produce the observed column densities.
The large size implies that the Hubble flow across an absorber
can be substantial.  Indeed, for a typical low column density line
in the simulations, the primary contribution to the line width
($b$-parameter) comes from Hubble flow.  This situation contrasts
with that in traditional conceptions of the forest, where lines
are assumed to be broadened by thermal motions of the gas or by
``turbulent'' motions of cloudlets.
Gravitationally induced peculiar velocities do distort the lines in
cosmological simulations, but these {\it coherent} flows are not
at all like Gaussian turbulence, where there is a large {\it dispersion}
in the velocities at a given spatial position.
Furthermore, because most lines with $\Nh \simgt 10^{14}\;\cdunits$ 
arise in moderately overdense regions that are expanding
slower than the Hubble rate, peculiar velocities on average have
the effect of {\it narrowing} \lya forest lines, not broadening them.

\begin{figure}[tb]
\centerline{\vbox{
\psfig{figure=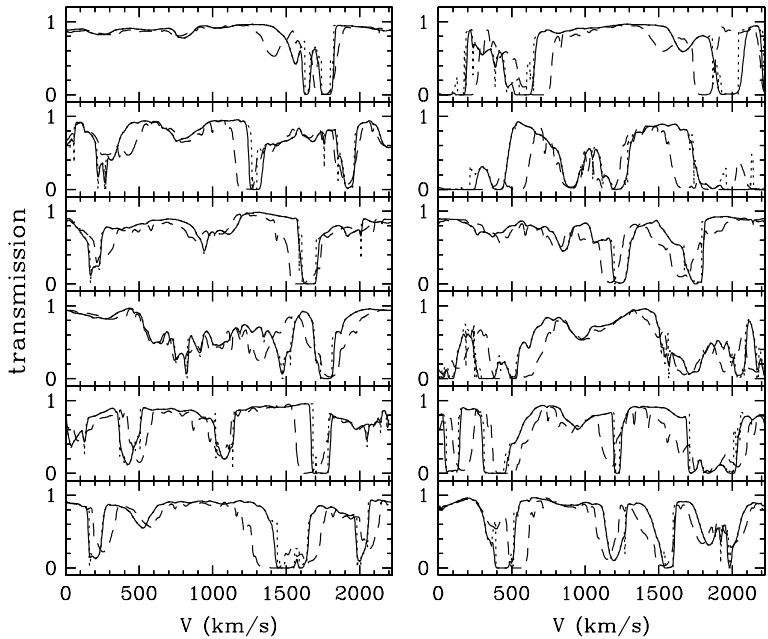,width=4.0truein}
}}
\caption[]{Solid lines show spectra along twelve randomly chosen lines of sight
through a simulation of the standard CDM model, at $z=3$.
Dotted and dashed lines show spectra along the same lines of sight
with no thermal broadening and no peculiar velocities, respectively.
}
\end{figure}

Figure 1 illustrates these points using spectra extracted along twelve
random lines of sight through a hydrodynamic
simulation of the ``standard'' CDM model
(SCDM, with $\Omega=1$, $h=0.5$, $\sigma_8=0.7$) at $z=3$
(see \cite{hkwm96} and \cite{cwkh97} for details).
Solid lines show the full \lya absorption spectra.
Dotted lines show the spectra with no thermal broadening --- they
are computed by artificially setting the gas temperature to zero 
(without changing the neutral fraction).  There are a handful of
sharp features in the non-thermally-broadened spectra that are
smoothed away in the full spectra.  However, in most regions 
the dotted and solid lines are barely distinguishable,
demonstrating that thermal broadening usually does not contribute
significantly to the width of the absorption features.
The dashed lines in Figure~1 show spectra with thermal broadening
but no peculiar motions --- they are extracted along the same lines
of sight after setting the peculiar velocities of all gas particles to zero.
Comparing the solid lines to the dashed lines shows that
peculiar velocities do shift the positions and distort the shapes
of individual absorption features, but they do not make them systematically
broader.  Indeed, as expected from the physical argument above, the
lines in the full spectra tend to be somewhat narrower than the lines
in the spectra with no peculiar velocities.

Studies of QSO pairs \cite{bechtold94,dinshaw94,dinshaw95} provide 
observational evidence for Hubble flow broadening of \lya forest lines
independently of cosmological simulations.
The inferred transverse coherence scale of the absorbers,
$l_t \sim 150 h^{-1}\;$kpc at $z \sim 2$, corresponds (for $\Omega=1$) 
to a line-of-sight extent $H(z)l_t \approx 80\;\kms$,  
considerably larger than the $\sim 25\;\kms$ $b$-parameter of
a typical forest line \cite{hu95}.
Nonspherical absorbers would be preferentially intercepted ``face on,''
but unless the absorbing structures are {\it highly} flattened
the observed transverse scale implies that Hubble flow across them
must make a major contribution to the line width.
The transverse coherence could in principle be a signature of clustering of 
small clouds rather than a physical scale of large absorbers, but the
nearly perfect coincidence of lines towards small-separation
gravitational lens pairs \cite{smette92,smette95} argues against
the clustering interpretation, and the detailed match of absorption
features shown by Rauch in these proceedings seems to rule it out
definitively.

Many papers have remarked on the low density of the absorbing
gas in cosmological simulations (e.g., 
\cite{cen94,zhang95,hkwm96,miralda96,zhang97}).  The issue of
Hubble flow broadening has received less attention, but
its implications are perhaps even more profound.
For a start, it means that the profile of a \lya forest line
shows a line-of-sight density profile through the absorbing
structure, albeit one that is non-linear and distorted by
peculiar motions.  This is not the case in the thermal broadening
picture, where the absorber itself is compact and the wings of
the line arise from high velocity atoms.  In the cosmological
picture, line wings show the absorbing structure itself fading
into the background, like mountains into foothills.

Another consequence of Hubble flow broadening is that the
Gunn-Peterson \cite{gunn65} formula,
\begin{equation}
\tau_{\rm GP} = \frac{\pi e^2}{m_e c}\; f_\alpha \lambda_\alpha
H^{-1}(z) \nh,
\label{taugp}
\end{equation}
provides a good approximation to the relation between local
\lya optical depth and the local space density of neutral hydrogen.
In a thermally broadened, compact cloud model, by contrast,
the optical depth is lower than $\tau_{\rm GP}$ at the line center
(the redshift space location of the dense cloud) and higher than
$\tau_{\rm GP}$ in the line wings (where there is no gas at the
corresponding redshift space position).
Cosmological simulations imply that the \lya forest can
be viewed as a fluctuating Gunn-Peterson effect, produced by
an inhomogeneous, diffuse intergalactic medium. 

What turns the fluctuating Gunn-Peterson idea from a novelty into
a powerful conceptual tool is the simplicity of the physics that
governs the ionization state of the low density gas.
This gas is in photoionization equilibrium, so the neutral hydrogen
density is $\nh \propto \rho^2 T^{-0.7}/\Gamma$, where $\Gamma$ is
the photoionization rate and the $T^{-0.7}$ factor accounts for the
temperature dependence of the hydrogen recombination coefficient near
$T \sim 10^4\;\K$.
The interplay between photoionization heating and adiabatic cooling
leads to a tight relation between temperature and density,
which can be well approximated by a power law, $T=T_0(\overden)^\gamma$
\cite{cwkh97,hg97}.  The values of $T_0$ and $\gamma$ depend on
the UV background spectrum and reionization history and can be 
computed semi-analytically \cite{hg97};
typically $T_0 \sim 6000\;$K and $\gamma \sim 0.3-0.6$.
With this physical reasoning, equation~(\ref{taugp}) can 
be converted to a formula we can describe as the
{\it fluctuating Gunn-Peterson approximation},
\begin{eqnarray}
\tau(\lambda_{\rm obs}) = &
 0.172 \left(\frac{\rho}{\overline \rho}\right)^\beta
\left(1 + \frac{dV_{\rm los}}{H(z) dx}\right)^{-1}
\left(\frac{1+z}{4}\right)^6
\left(\frac{H(z)/H_0}{5.51}\right)^{-1} h^{-1} \;\times & \nonumber \\
& \left(\frac{\Omega_b h^2}{0.0125}\right)^2 
\left(\frac{T_0}{10^4\;{\rm K}}\right)^{-0.7}
\left(\frac{\Gamma}{10^{-12}\;{\rm s}^{-1}}\right)^{-1}\;, & \label{fgpa} 
\end{eqnarray}
where $\beta \equiv 2-0.7\gamma$,
$\overden$ is the overdensity at the position where the
redshift (cosmological plus peculiar velocity) is 
$\lambda_{\rm obs}/\lambda_\alpha - 1$, and $dV_{\rm los}/dx$
is the derivative of the line-of-sight peculiar velocity at the same
position.  The peculiar velocity term accounts for the mapping
from real space to redshift space.  
In principle, $\overden$ here refers to the {\it gas} overdensity,
but because the temperature is low, pressure gradients are small
compared to gravitational forces, and the gas traces the dark matter
quite well.

Equation~(\ref{fgpa}) is valid if all gas lies on the temperature-density
relation and thermal broadening and collisional ionization can be
ignored.  The approximation breaks down when $\overden \ga 10$,
but these regions occupy a small fraction of the spectrum.
They are responsible for high column density lines, and the general
description presented above begins to break down for
$\Nh \simgt 10^{15}-10^{16}\;\cdunits$.  This description might
also become less accurate at low redshifts; 
we have not examined the \lya forest
at $z<2$ with our simulations.  We should also note that
{\it some} low column density lines at high $z$
arise in shock heated gas and are thermally broadened.

A second useful approximation arises from ignoring peculiar velocities,
setting $dV_{\rm los}/dx=0$ in equation~(\ref{fgpa}), so that there is
a one-to-one relation between optical depth and overdensity.
Figure~6 of \cite{cwkh97} shows that the $\tau-\rho$ relation remains
tight in simulated spectra even when peculiar velocities, thermal
broadening, shock heating, and collisional ionization are all taken
into account.  The distribution of \lya optical depths $P(\tau)$
is directly observable from high resolution QSO spectra \cite{rauch97},
and one can use this second approximation to write the mean density
of the ``warm'' IGM that produces the \lya forest in terms of an
integral over this distribution,
${\overline\rho}_{\rm WIGM} = \int_0^\infty \rho(\tau) P(\tau)d\tau$.
After some manipulation, one obtains the density parameter of this 
warm IGM component \cite{wkh97,wmhk97},
\begin{eqnarray}
\Omega_{\rm WIGM} 
= & 0.021 h^{-3/2}
\left(\frac{
\left[\int_0^\infty \tau^{1/\beta} P(\tau) d\tau\right]^{\beta/2}}
{0.70} \right) \left(\frac{4}{1+z}\right)^3 \;\times & \nonumber \\
& \left(\frac{H(z)/H_0}{5.51}\right)^{1/2}
\left(\frac{T_0}{10^4\;\K}\right)^{0.35}
\left(\frac{\Gamma}{10^{-12}\; {\rm s}^{-1}}\right)^{1/2}\;. & \label{igm}
\end{eqnarray}
For the fiducial value of the optical depth integral, we have used
a value inferred from the observations of \cite{rauch97} at $z=3$.
The implied $\Omega_{\rm WIGM}$ is a substantial fraction of the
baryon density parameter $\Omega_b$ allowed by big bang nucleosynthesis,
indicating that most of the baryons in the universe at $z \sim 3$
resided in the \lya forest, as the simulations predict.

The relation between the underlying mass distribution and the number density
of \lya forest lines in a spectrum may be physically complex, and it
is sensitive to the details of the observational procedures and the
method used deblend absorption features.
However, the fluctuating Gunn-Peterson approximation implies that
the relation between mass density and observed flux is direct and simple.
If one wants to use the \lya forest to test theories of structure 
formation, it is best to abandon 
lines altogether and treat the full observed spectrum as a continuous field.
Statistical properties of this ``flux field'' are directly related to
statistical properties of the underlying density and velocity fields,
which are basic predictions of cosmological models.
We are currently studying a variety of statistical measures similar to
those used in large scale structure analyses, applying them to simulations
in order to assess their sensitivity to different properties of the
initial fluctuations and to values of cosmological parameters.
We have also developed and tested a method to recover the shape and
amplitude of the primordial mass power spectrum $P(k)$ from \lya forest data 
\cite{cwkh97b}, again motivated by the ``continuous field'' point of view.
The 3-d flux power spectrum has the same shape as the mass power spectrum
on large scales, and the normalization can be determined by evolving numerical 
simulations with this initial $P(k)$ shape until they reproduce the 
observed power spectrum of the QSO flux. Imposing the
observed mean \lya opacity as a constraint makes the derived $P(k)$ 
normalization insensitive to the choice of cosmological parameters, 
ionizing background spectrum, or reionization history. 
This approach thus neatly circumvents the uncertain physics of galaxy
formation and ``biasing,'' which complicates the interpretation of
power spectra measured from galaxy redshift surveys.
Application to existing samples of QSO spectra should soon yield
the power spectrum of mass fluctuations in the high redshift universe.

\vfill

\begin{iapbib}{99}{
\bibitem{bechtold94}
Bechtold J., Crotts A.P.S., Duncan R.C., Fang Y., 1994, \apj 437, L83
\bibitem{bi93}
Bi H.G., 1993, \apj 405, 479
\bibitem{cen94}
Cen R., Miralda-Escud\'e J., Ostriker J.P., Rauch M., 1994, \apj 437, L9
\bibitem{cwkh97}
Croft R.A.C., Weinberg D.H., Katz N., Hernquist L., 1997, \apj in press
(astro-ph/9611053)
\bibitem{cwkh97b}
Croft R.A.C., Weinberg D.H., Katz N., Hernquist L., 1997,
\apj submitted (astro-ph/9708018)
\bibitem{dinshaw94}
Dinshaw N., Impey C.D., Foltz C. B., Weymann R.J., 
Chaffee F.H., 1994, \apj 437, L87
\bibitem{dinshaw95}
Dinshaw N., Foltz C.B., Impey C.D., Weymann R.J., Morris S.L., 1995, 
Nature 373, 223
\bibitem{gunn65}
Gunn J.E., Peterson B.A., 1965, \apj 142, 1633
\bibitem{hkwm96}
Hernquist L., Katz N., Weinberg D.H.,
Miralda-Escud\'e J. 1996, \apj 457, L5
\bibitem{hu95}
Hu E.M., Kim T.S., Cowie L.L., Songaila A., Rauch M., 1995, \aj
110, 1526
\bibitem{hg97}
Hui L., Gnedin N., 1997, \mn submitted (astro-ph/9612232)
\bibitem{miralda96}
Miralda-Escud\'e J., Cen R., Ostriker J.P., Rauch M., 1996, \apj 471, 582
\bibitem{rauch97}
Rauch M., Miralda-Escud\'e J., Sargent W.L.W., Barlow T.A.,
Weinberg D.H., Hernquist L., Katz N., Cen R., Ostriker J.P.,
1997, \apj in press (astro-ph/9612245)
\bibitem{smette92}
Smette A., Surdej J., Shaver P.A., Foltz C.B., Chaffee F.H.,
Weymann R.J., Williams R.E., Magain P., 1992, \apj 389, 39
\bibitem{smette95}
Smette A., Robertson J.G., Shaver P.A., 
Wisotzki, L., Koehler, T., 1995, A\&AS, 113, 199
\bibitem{wkh97}
Weinberg D.H., Katz N., Hernquist L., 1997,
in Origins, eds. J. M. Shull, C. E. Woodward, \& H. Thronson,
(ASP Conference Series: San Francisco), (astro-ph/9708213)
\bibitem{wmhk97}
Weinberg D.H., Miralda-Escud\'{e} J., Hernquist L., Katz N., 1997,
\apj 490, in press (astro-ph 9701012)
\bibitem{zhang95}
Zhang Y., Anninos P., Norman M.L., 1995, \apj 453, L57
\bibitem{zhang97}
Zhang Y., Meiksin A., Anninos P., Norman M.L., 1997, \apj in press
}
\end{iapbib}
\end{document}